\documentclass[conference]{IEEEtran}
\IEEEoverridecommandlockouts

\usepackage{booktabs}
\usepackage{multirow} 
\usepackage{amsmath,amssymb,amsfonts}
\usepackage{algorithmic}
\usepackage{graphicx}
\usepackage{xcolor}
\usepackage{xurl}
\usepackage[retain-explicit-plus]{siunitx}
\def\BibTeX{{\rm B\kern-.05em{\sc i\kern-.025em b}\kern-.08em
 T\kern-.1667em\lower.7ex\hbox{E}\kern-.125emX}}

\usepackage{listings}
\usepackage{amssymb}
\usepackage{pifont}
\newcommand{\cmark}{\ding{51}}
\newcommand{\xmark}{\ding{55}}

\begin{document}

\title{Promoting security and trust on social networks: Explainable cyberbullying detection using Large Language Models in a stream-based Machine Learning framework}

\author{\IEEEauthorblockN{Silvia García-Méndez}
\IEEEauthorblockA{\textit{Information Technologies Group - atlanTTic} \\
\textit{University of Vigo}\\
Vigo, Spain \\
sgarcia@gti.uvigo.es | 0000-0003-0533-1303}
\and
\IEEEauthorblockN{Francisco de Arriba-Pérez}
\IEEEauthorblockA{\textit{Information Technologies Group - atlanTTic} \\
\textit{University of Vigo}\\
Vigo, Spain \\
farriba@gti.uvigo.es | 0000-0002-1140-679X}
}

\maketitle

\textcolor{teal}{© 2024 IEEE. Personal use of this material is permitted. Permission from IEEE must be obtained for all other uses, in any current or future media, including reprinting/republishing this material for advertising or promotional purposes, creating new collective works, for resale or redistribution to servers or lists, or reuse of any copyrighted component of this work in other works.}\\

\begin{abstract}
Social media platforms enable instant and ubiquitous connectivity and are essential to social interaction and communication in our technological society. Apart from its advantages, these platforms have given rise to negative behaviors in the online community, the so-called cyberbullying. Despite the many works involving generative Artificial Intelligence (\textsc{ai}) in the literature lately, there remain opportunities to study its performance apart from zero/few-shot learning strategies. Accordingly, we propose an innovative and real-time solution for cyberbullying detection that leverages stream-based Machine Learning (\textsc{ml}) models able to process the incoming samples incrementally and Large Language Models (\textsc{llm}s) for feature engineering to address the evolving nature of abusive and hate speech online. An explainability dashboard is provided to promote the system's trustworthiness, reliability, and accountability. Results on experimental data report promising performance close to \SI{90}{\percent} in all evaluation metrics and surpassing those obtained by competing works in the literature. Ultimately, our proposal contributes to the safety of online communities by timely detecting abusive behavior to prevent long-lasting harassment and reduce the negative consequences in society.
\end{abstract}


\section{Introduction}

Online communication has become an essential feature of social interaction thanks to the proliferation of new communities and networks \cite{islam2023comparative}. More in detail, 4.90 billion people reported using social media worldwide in 2023, and this figure is expected to rise to 5.85 billion by 2027\footnote{Available at \url{https://www.forbes.com/advisor/business/social-media-statistics}, October 2024.}. Even though these platforms enable instant, ubiquitous connectivity, they have given rise to negative behaviors such as bullying, racism, sexism, and trolling \cite{Saifullah2024}. Regarding cyberbullying statistics by the World Health Organization in Europe\footnote{Available at \url{https://www.who.int/europe/news/item/27-03-2024-one-in-six-school-aged-children-experiences-cyberbullying--finds-new-who-europe-study}, October 2024.}, about \SI{12}{\percent} adolescents report harassing others. In contrast, \SI{15}{\percent} have experienced abuse related to sharing sensitive or private content, receiving threatening comments and messages, or being the target of false information and rumors spreading.

The consequences of cyberbullying range from low self-esteem and increased negative emotional state, depression, self-harm, and changes in sleeping patterns to suicidal ideation \cite{Arisanty2022,Verma2022,Saini2023}. The use of abusive and hateful language, even if being a one-off by isolated users, can trigger repetitive behavior in the online community and, unfortunately, result in the so-mentioned cyberbullying \cite{Herodotou2020}. Thus, it has a direct detrimental impact on users' mental health \cite{islam2023comparative,vanpech2024detecting}.

To address this appalling situation, social media companies have followed reporting strategies. It is the case of \textsc{x}\footnote{Available at \url{https://x.com}, October 2024.} with the \textsc{dcma} claim, which can lead to account suspension or delegation, similar to what TikTok\footnote{Available at \url{https://www.tiktok.com}, October 2024.} is doing. Moreover, Instagram\footnote{Available at \url{https://www.instagram.com}, October 2024.} tests for negative and inappropriate comments and messages with its \textit{Hidden Words} feature. Unfortunately, this results from the recent spike of online harassment or cyberbullying across these popular platforms \cite{Herodotou2020}. However, the mentioned user-dependent prevention mechanisms (\textit{e.g.}, blocking, removing, and reporting) may become ineffective due to their passive and manual nature \cite{Teng2023}.

Of interest to social media data analytics, consideration should be given to the continuous pace of social media content, \textit{i.e.}, in a stream-based manner. At the same time, how online users express themselves is permanently evolving. The latter is especially relevant to analyzing aggressive behavior in social media properly \cite{Herodotou2020}. Given this prevalent concern in the digital age, cyberbullying detection has attracted the attention of academia \cite{islam2023comparative}. Note that detecting cyberbullying is a complex, challenging task in which the use of 
Artificial Intelligence (\textsc{ai}) approaches such as Machine Learning (\textsc{ml}) models may not be sufficient \cite{Cheng2021,Muneer2023}. 

In this regard, Large Language Models (\textsc{llm}s) represent an emerging approach with promising opportunities for Natural Language Processing (\textsc{nlp}). However, their leveraging with other technologies like \textsc{ml} or image recognition is still in its infancy in many fields, such as cyberbullying detection \cite{vanpech2024detecting}. More in detail, \textsc{llm}s such as those developed by Open\textsc{ai}\footnote{Available at \url{https://openai.com}, October 2024.} (\textit{e.g.}, \textsc{gpt-4}) have the capabilities to process and generate human-like text thanks to the underlying neural network architecture composed of millions of parameters across hundreds of layers \cite{Daniel2023}. Since they are trained with diverse Internet data, these models can learn to detect various linguistic patterns and contextual insights \cite{vanpech2024detecting}.

Despite the many works involving generative \textsc{ai} in the literature lately, there remain opportunities to study its performance apart from being final solutions, \textit{i.e.}, asking the models to classify social media content. Conversely, we propose to leverage them to extract high-level reasoning features that can be exploited by an \textsc{ml} model with explainable capabilities for cyberbullying detection. Moreover, note that while traditional \textsc{ml} models benefit from the textual pattern analyzing capabilities of \textsc{llms}, the latter generative \textsc{ai} may have limited contextual understanding of the classification problem, which we plan to circumvent with supervised \textsc{ml}.

Unfortunately, the \textsc{ai} technologies for data analytics, including cyberbullying detection through text mining in the literature, are opaque, which not only affects the intrinsic interpretability and trustworthiness of the algorithms and reliability and accountability of the results provided but also the acceptance and adoption of these solutions by social media platforms, policymakers, and online users \cite{Tesfagergish2024}. Accordingly, eXplainable Artificial Intelligence (\textsc{xai}) seems appropriate to describe the functioning of autonomous machine-based decision models, which mainly operate as black boxes for the general public and developers \cite{Gongane2023}. Among the \textsc{xai} models, consideration should be given to Deep\textsc{lift}, layer-wise relevance propagation, \textsc{lime} \cite{Mehta2022}. In short, \textsc{xai} techniques promote interoperability, transparency, and fairness regarding why a particular message was flagged \cite{Tesfagergish2024}.

Ultimately, timely detection of abusive behavior on social networks is essential to prevent long-lasting harassment and reduce the negative consequences in the community \cite{Teng2023}. Summing up, we propose an innovative and real-time solution for cyberbullying detection that leverages stream-based \textsc{ml} and \textsc{llm}s for feature engineering to address the evolving nature of abusive and hate speech online. More in detail, it can process samples incrementally, enhancing the model's adaptability. Our work's contributions are completed with explainability capabilities to promote trust in \textsc{ai} among society.

The remainder of this manuscript is divided into the following sections. Section \ref{sec:related_work} summarizes the central literature on cyberbullying detection. Section \ref{sec:methodology} details our system architecture, while Section \ref{sec:results} shows the results obtained with our methodology and compares them with other works in the literature. Finally, Section \ref{sec:conclusion} does the main conclusions of this work and proposes future analyses.

\section{Literature review}
\label{sec:related_work}

The most straightforward approach to cyberbullying detection is keyword-based filtering \cite{Tesfagergish2024}. Then, early research focused on \textsc{ml} classifiers \cite{Balkir2022} and, subsequently, deep learning models, especially Convolutional Neural Networks (\textsc{cnn}) and Recurrent Neural Networks (\textsc{rnn}), \cite{Paul2022,Mazari2023}. More in detail, the traditional \textsc{ml} models leverage linguistic patterns from the textual content and social network structures \cite{islam2023comparative,vanpech2024detecting}. In this regard, the literature endorses the superior performance of the former traditional \textsc{ml} architectures when considering both efficiency and effectiveness \cite{Chatzakou2019}. The main concern related to the deep learning frameworks lies in their intensive requirements regarding computational and data resources \cite{Tesfagergish2024}. 

Most works in the literature compared the performance of \textsc{llm}s with baseline \textsc{ml} models and deep learning. It is the case of S. Paul and S. Saha (2022) \cite{Paul2022} who investigated the performance of a finetuned version of \textsc{bert} to detect cyberbullying employing data from Twitter, Wikipedia, and Formspring. Additionally, K. Verma \textit{et al.} (2022) \cite{Verma2022} compared traditional \textsc{ml} models (\textit{i.e.}, Support Vector Machines) with \textsc{llm}s (\textit{e.g.}, \textsc{bert}). Regarding deep learning approaches, A. Muneer \textit{et al.} (2023) \cite{Muneer2023} analyzed the performance of a finetuned \textsc{bert} model. Ultimately, similar to the results obtained by S. Paul and S. Saha (2022) \cite{Paul2022}, K. Verma \textit{et al.} (2022) \cite{Verma2022} and A. Muneer \textit{et al.} (2023) \cite{Muneer2023}, in the more recent works by T. H. Teng and K. D. Varathan (2023) \cite{Teng2023} and K. Saifullah \textit{et al.} (2024) \cite{Saifullah2024} the performance of both \textsc{llm}s and \textsc{ml} models were comparable which inspired us to leverage its combination.

Moreover, among those solutions in the literature that address cyberbullying detection combining \textsc{llm}s and \textsc{ml}, Y. Yan \textit{et al.} (2023) \cite{Yan2023} used Chat\textsc{gpt} for global knowledge extraction to feed the K-Nearest Neighbor (\textsc{knn}) algorithm. Unlike in our proposal, no high-level reasoning knowledge is extracted. In fact, prior works in state of the art demonstrate that simply relying on text-derived features limits the performance of the solutions due to the diversity and complexity of speech (\textit{e.g.}, irony, sarcasm) \cite{Wich2021}. Accordingly, consideration should be given to the difficulty of extracting nuanced contextual subtleties from word embeddings \cite{islam2023comparative}. As mentioned, we prevent this issue by exploiting \textsc{llm}s to extract expert users' features to train the \textsc{ml} models. Mainly, the use of \textsc{llm}s reduces the limitations of prior research regarding the reliance on hand-crafted features and the low generalizability due to the reduced capability of traditional \textsc{nlp} strategies (\textit{e.g.}, bag of words) to model the variety of language \cite{Muneer2023}.

\begin{table*}[!htbp]
\centering
\caption{Comparison of related solutions.}
\label{tab:comparison}
\begin{tabular}{lccccc}
\toprule
\textbf{Authorship} & \textbf{Approach} & \bf Operation & \textbf{Explainability}\\
\midrule

H. Herodotou et al. (2020) \cite{Herodotou2020} & \textsc{ml} & Streaming & \xmark\\

M.S. Islam and R. I. Rafiq (2023) \cite{islam2023comparative} & \textsc{llm}s & Batch & \xmark\\

A. Sadek et al. (2023) \cite{Sadek2023} & \textsc{llm}s & Batch & \xmark\\

Y. Yan \textit{et al.} (2023) \cite{Yan2023} & \textsc{ml} + \textsc{llm}s & Batch & \xmark\\

E. A. Nina-Gutiérrez et al. (2024) \cite{nina2024multilingual} & \textsc{llm}s & Batch & \xmark\\

\midrule

Our proposal & \textsc{ml} + \textsc{llm}s & Streaming & \cmark\\
 
\bottomrule
\end{tabular}
\end{table*}

Note that all the aforementioned approaches to cyberbullying detection are deployed on a batch basis, which impacts the capacity of the models to adapt over time and limits the continuous monitoring of social media content \cite{Rafiq2018}. The sole exception to the best of our knowledge is the work by H. Herodotou \textit{et al.} (2020) \cite{Herodotou2020}. The authors presented a real-time aggression detection solution for Twitter by exploiting stream-based machine learning models. The authors used the Hoeffding Tree, Adaptive Random Forest, and Streaming Logistic Regression models. However, instead of using \textsc{llms} for feature extraction, they followed the bag-of-words approach. Conversely, we propose a stream-based \textsc{ml} framework able to process the incoming samples incrementally and leverage \textsc{llm}s to address the evolving nature of online hate speech and abusive language. Note that the system's ability to be understood by end users thanks to the explainability module is another relevant contribution of our research.

When it comes to the appropriateness of generative \textsc{ai} to this task, M.S. Islam and R. I. Rafiq (2023) \cite{islam2023comparative} studied the effectiveness of \textsc{llm}s for cyberbullying detection on Instagram. For that purpose, the authors explored different models (\textit{i.e.}, Chat\textsc{gpt} 3.5 Turbo and Text-davinci-003), learning strategies (\textit{i.e.}, zero-shot and one-shot) and prompts. Special mention deserves the work by A. Sadek \textit{et al.} (2023) \cite{Sadek2023} who studied the performance of traditional \textsc{ml} models, and \textsc{llm}s (\textit{e.g.}, \textsc{bert} and Chat\textsc{gpt}) for Arabic cyberbullying. However, the results show the limited performance of models like Chat\textsc{GPT} to solve classification problems on their own. Nevertheless, as we did in our research, they can be leveraged with other technologies like standard \textsc{ml} models. Moreover, E. A. Nina-Gutiérrez \textit{et al.} (2024) \cite{nina2024multilingual} used a finetuned version of \textsc{gpt-3.5} to detect hate speech and offensive language. As in the work by A. Sadek \textit{et al.} (2023) \cite{Sadek2023}, the authors did not explore the use of \textsc{llm}s beyond its direct application as final decision makers with limited contextual understanding of the classification problem.

Ultimately, regarding the use of interpretable models or explainability techniques, we must mention the early work by M. Wich \textit{et al.} (2021) \cite{Wich2021}. They assessed the vulnerability of their \textsc{bert}-based solution for abusive language classification with SHapley Additive exPlanations (\textsc{shap}). Similarly, T. Ahmed \textit{et al.} (2022) \cite{ahmed2022performance} used \textsc{gpt-2} among other \textsc{llms} (\textit{e.g.}, \textsc{bert}) to extract the embeddings to detect cyberbullying through ensemble learning. Regarding interpretability, just word importance indices are provided for the \textsc{bert} model. Another representative example of the use of \textsc{shap} is the research by P. Aggarwal and R. Mahajan (2023) \cite{aggarwal2024shielding} who combined \textsc{bert} and Supper Vector Machines (\textsc{svm}) for multi-class cyberbullying classification. Additionally, both H. Mehta and K. Passi \cite{Mehta2022} and V. Pawar \textit{et al.} (2022) \cite{Pawar2022} exploited Local Interpretable Model-Agnostic Explanations (\textsc{lime}) to provide interpretable data for hate speech detection, similar to the approach by Shakil and M. G. R. Alam \cite{shakil2022hate}. Furthermore, V. U. Gongan \textit{et al.} (2023) \cite{Gongane2023} compared the performance of neural network models (\textit{i.e.}, \textsc{cnn}, Bi\textsc{lstm} and \textsc{bert}) and leveraged their results with \textsc{lime}. Ultimately, M. Umer \textit{et al.} (2024) \cite{umer2024cyberbullying} combined the Ro\textsc{bert}a model with the principle component analysis (\textsc{pca}) and global vectors for word representation (\textsc{glove}) to extract word embedding features. Then, the authors provided interpretable results using \textsc{lime}. Unfortunately, the latter explainable techniques (\textit{i.e.}, \textsc{shap} and \textsc{lime}) only provide the probability score of the features that motivated the model's decision, which continues to be challenging for the end users to comprehend.

\subsection{Research contributions}

Table \ref{tab:comparison} shows the most closely related solutions to easily compare and assess our contributions. As described in the literature review, most works perform cyberbullying detection on a batch basis except for H. Herodotou et al. (2020) \cite{Herodotou2020}. However, this work relies exclusively on the bag-of-words approach for text mining, which is very limited in its ability to model the variety of language. Additionally, Y. Yan \textit{et al.} (2023) \cite{Yan2023} is the only work in which \textsc{llm}s are exploited for feature engineering, a new contribution of our work since the latter is exclusively used to extract global knowledge (\textit{i.e.}, the \textsc{llm}s are asked to summarize the main aspects in the post). Moreover, no explainability is provided in any of the selected works. 

Summing up, the main contributions of the proposed solution for the cyberbullying detection field are as follows:
\begin{itemize}
 \item A stream-based framework that processes samples incrementally, enhancing the model's adaptability.
 \item The use of \textsc{llm}s to extract high-level reasoning features used to train the \textsc{ml} models.
 \item The explainability dashboard which promotes trustworthiness, reliability, and accountability.
\end{itemize}

\section{Methodology}
\label{sec:methodology}

\begin{figure}[!htpb]
 \centering
 \includegraphics[scale=0.1]{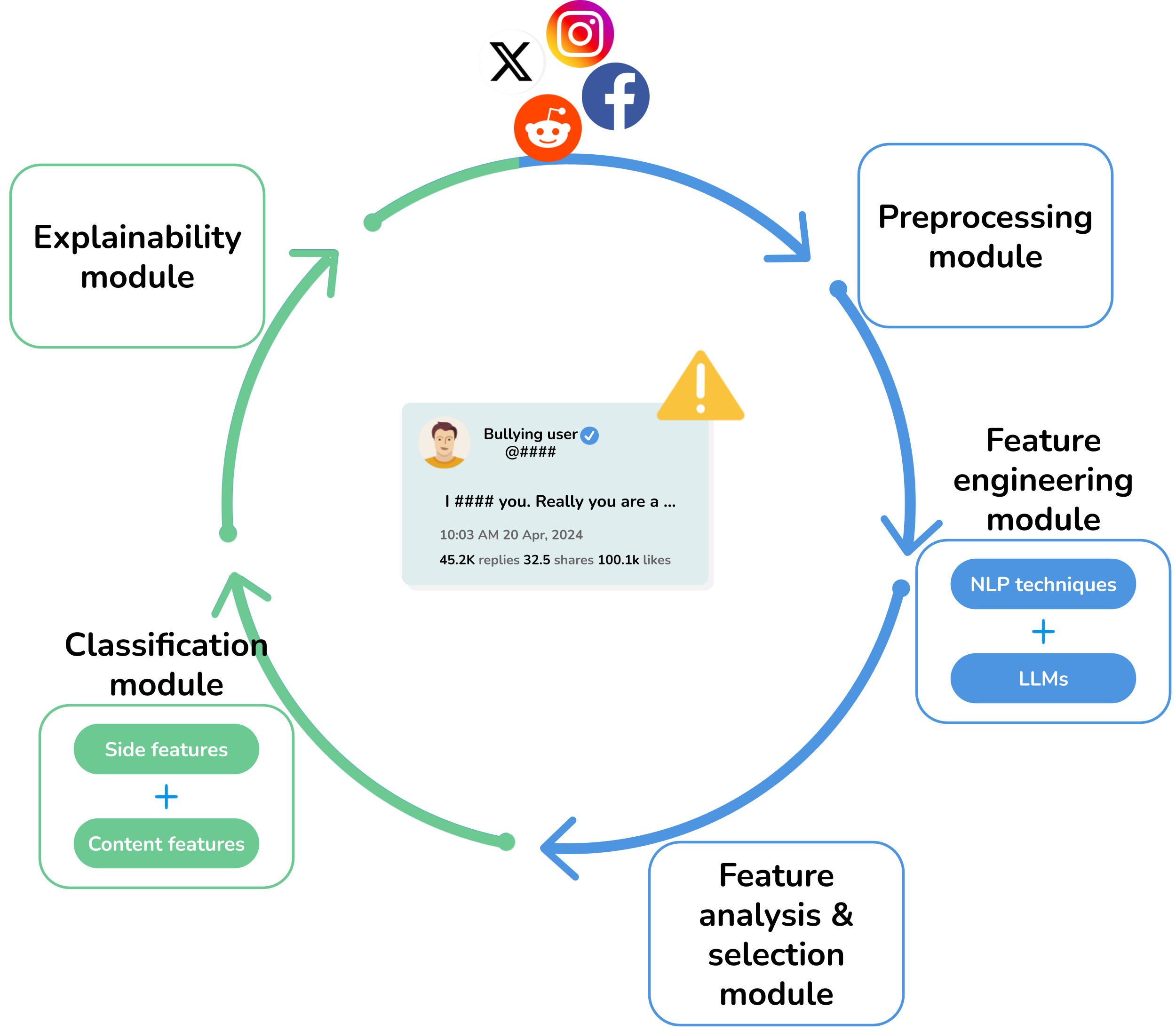}
 \caption{System scheme.}
 \label{fig:scheme}
\end{figure}

Figure \ref{fig:scheme} shows the architecture of the proposed solution composed of (\textit{i}) a preprocessing module to remove unmeaningful data from the social networks posts; (\textit{ii}) a feature engineering module to generate new features by combining traditional \textsc{nlp} and prompt engineering techniques to take the most of the language modeling capabilities of \textsc{llm}s; (\textit{iii}) feature analysis and selection module, where the relevant data are extracted to train the \textsc{ml} models in (\textit{iv}) the classification module; and (\textit{v}) the explainability module to generate natural language descriptions of the predictions provided by the system.

Moreover, two stages exist in the system: (\textit{i}) the cold start (see the blue color representation of this stage in Figure \ref{fig:scheme}) and (\textit{ii}) the streaming setting. More in detail, in the initial cold start stage, the first samples (see Section \ref{sec:feature_analysis_selection} and Section \ref{sec:feature_analysis_selection_results}) are used to optimize the hyperparameter of the classifiers and select the initial set of relevant features, which minimizes the computational load in the streaming operation. When this stage ends, the system maintains a constant data flow involving all modules in Figure \ref{fig:scheme}.

\subsection{Preprocessing module}
\label{sec:preprocessing}

The experimental data composed of social media posts are processed before feature engineering with \textsc{nlp} techniques (see Section \ref{sec:feature_engineering} for technical information). Notably, using regular expressions, the system removes numbers, punctuation characters, links, blank spaces, two-character words, and special characters. Moreover, the stop words are removed, and the content is tokenized and lemmatized. Finally, isolated characters are also removed.

\subsection{Feature engineering module}
\label{sec:feature_engineering}

Table \ref{tab:features} lists all the features used in the system. Features 1-7 are calculated using prompt engineering applied to an \textsc{llm} model, resulting in boolean information (see Section \ref{sec:feature_analysis_selection_results} for implementation details). The text sent for analysis to the model is not preprocessed and includes all the original raw information. In addition, features 8-16 are calculated by applying traditional \textsc{nlp} techniques on the preprocessed content (see Section \ref{sec:preprocessing}).

While features 1-7 are remarks that characterize cyberbullying posts and allow for interpreting the interlocutor's intentionality, features 8-15 are related to language characteristics. Both are considered side features, independent of the content and therefore much more flexible. Our study compares the latter with the content-dependent feature 16 derived from the word-matrix method.

\begin{table*}[!htbp]
\centering
\caption{\label{tab:features}Features engineered for cyberbullying detection.}
\begin{tabular}{llllp{6cm}}
\toprule
\bf Technique & \bf Category & \bf ID & \bf Name & \bf Description\\\hline

\multirow{10}{*}{\textsc{llm}} 
& \multirow{10}{*}{Side} 
& 1 & Derogatory remarks & It indicates if the post contains insults.\\
& & 2 & Humiliating language & It indicates if the post contains humiliating and depersonalizing expressions.\\
& & 3 & Racist terms & It indicates explicit references to racist terms.\\
& & 4 & Sarcasm terms & It indicates if the post is sarcastic.\\
& & 5 & Sexual terms & It indicates if the post contains sexual terms.\\
& & 6 & Threatening remarks & It indicates if the post contains threatening expressions.\\
& & 7 & Violence terms & It indicates if the post contains terms that promote violence.\\
\midrule

\multirow{9}{*}{\textsc{nlp}}
& \multirow{9}{*}{Side} 
& 8 & Difficult word counter & It counts complex terms.\\
&& 9 & Emotion load & Level of anger, fear, happiness, sadness, and surprise.\\
& & 10 & Flesch score & Readability level of the post.\\
& & 11 & McAlpine score & Non-native English speakers redability score.\\
& & 12 & Polarity & Negative, neutral, or positive load.\\
& & 13 & \textsc{pos} counters & Ratio of adjectives, determiners, nouns, pronouns, punctuation marks, and verbs vs. total words.\\
& & 14 & Reading time & An estimation of the post reading time.\\
& & 15 & Word count & Number of words in the post.\\

\cmidrule(lr){2-5}
& Content & 16 & Word-grams & Frequencies of word n-grams in the post.\\

\bottomrule
\end{tabular}
\end{table*}

\subsection{Feature analysis \& selection module}
\label{sec:feature_analysis_selection}

Streaming systems require high computational capacity because, at each interaction, the model is adapted to the newly labeled input samples. Therefore, this approximation cannot be evaluated with as many features as in batch analysis, making the classification problem even more challenging. For this reason, our system combines two methods for feature selection. The first one is applied during cold start and uses a meta-transformer wrapper \textsc{ml} model. This classifier selects the most relevant features based on a Mean Decrease in Impurity (\textsc{mdi}) technique \cite{Burkart2021}. The second filter is applied after the cold start stage ends and seeks to eliminate features with lower variance. This reduces the irrelevant features and the model size, resulting in improved performance. For our analyses, we use the default parameters of both selectors, which are further described in Section \ref{sec:feature_analysis_selection_results}.

\subsection{Classification module}
\label{sec:classification}

Two evaluation scenarios exist in our experiments.

\begin{itemize}

\item \textbf{Scenario 1} evaluates the \textsc{ml} models using only the side features (1-15 in Table \ref{tab:features}). It studies the quality of the results based on the appropriateness of prompt engineering and \textsc{nlp} to extract relevant data from the posts, independently of its explicit content (\textit{i.e.}, the word embeddings in feature 16).

\item \textbf{Scenario 2} evaluates the impact of content-dependent features (16 in Table \ref{tab:features}) in the classifiers evaluation metrics. Although these features are widely used in the literature \cite{ahmed2022performance,umer2024cyberbullying}, they are highly time-consuming regardless of the classification approach (\textit{i.e.}, in batch or streaming basis) and reduce the generalization of the solution to different experimental data, particularly relevant in highly variant layouts like social media knowledge mining. Moreover, this kind of data has been reported to fail to capture nuanced contextual subtleties \cite{islam2023comparative}.

\end{itemize}

Furthermore, to evaluate the solution, the following classifiers have been selected:

\begin{itemize}
 \item \textbf{Gaussian Naive Bayes} (\textsc{gnb}) \cite{Xu2018} which uses the Gaussian probability to minimize the prediction error.
 
 \item \textbf{Hoeffding Adaptive Tree Classifier} (\textsc{hatc}) \cite{Weinberg2023} is the streaming equivalent of the Decision Tree (\textsc{dt}) in batch setting. The prediction is computed relying on feature value data.
 
 \item \textbf{Adaptive Random Forest Classifier} (\textsc{arfc}) \cite{Zhang2021} is the streaming equivalent of the Random Forest (\textsc{rf}) in batch operation. This model comprises a configurable number of \textsc{dt} models, and the prediction is obtained based on majority voting.
\end{itemize}

\subsection{Explainability module}
\label{sec:explainability}

The natural language explanations of each prediction are generated using \textsc{llm}s leveraged with prompt engineering. The 1-7 side features in Table \ref{tab:features} are high-level reasoning features that capture users’ expert information by detecting linguistic patterns, language usage, and deep semantic meaning, taking advantage of the understanding capabilities of these models. In the end, the natural language explanation generated by the \textsc{llm} derives from the feature values, the post content itself, and the \textsc{ml} model prediction, along with its confidence to generate a holistic explanation of the detection of cyberbullying.

\section{Results and discussion}
\label{sec:results}

This section shows the results obtained for both scenarios under evaluation exploiting (\textit{i}) side and (\textit{ii}) content-dependent features, respectively. The analysis was performed on a computer with the following specifications:

\begin{itemize}
 \item \textbf{Operating System (\textsc{os})}: Ubuntu 18.04.2 LTS 64 bits
 \item \textbf{Processor}: Intel\@Core i9-10900K \SI{2.80}{\giga\hertz}
 \item \textbf{RAM}: \SI{96}{\giga\byte} DDR4 
 \item \textbf{Disk storage}: \SI{480}{\giga\byte} NVME + \SI{500}{\giga\byte} SSD
\end{itemize}

\subsection{Experimental data}

The dataset used\footnote{Available at \url{https://www.kaggle.com/datasets/andrewmvd/cyberbullying-classification}, October 2024.} \cite{Wang2020} is composed of social media posts extracted from \textsc{x} and labelled according to cyberbullying presence or absence. Table \ref{tab:dataset_distribution} summarizes its statistics\footnote{Compared to the original data, empty entries were removed, and the classes were balanced based on the number of samples for the majority class using the \texttt{RandomUnderSampler} library available at \url{https://imbalanced-learn.org/stable/references/generated/imblearn.under_sampling.RandomUnderSampler.html}, October 2024.}. 

The samples comprise 21.25$\pm$14.19 words on average, which are reduced to 9.64$\pm$7.14 when the preprocessing is performed. The latter improves the performance of the \textsc{ml} by reducing the amount of textual input data (\textit{i.e.}, it limits the size of feature 16 in Table \ref{tab:features} and the amount of time invested in computing features 8-15). Note that this reduction does not imply worsening the solution performance since the full raw content is analyzed by the \textsc{llm} (features 1-7). Hence, the advantage of combining both feature engineering methods.

\begin{table}[!htpb]\centering
\caption{Data set distribution.}
\label{tab:dataset_distribution}
\begin{tabular}{lc}\toprule
\textbf{Cyberbullying} & \multicolumn{1}{l}{\textbf{Samples}} \\
\midrule
Absent & 7945 \\
Present & 7945 \\ 
\midrule
Total & 15890 \\ 
\bottomrule
\end{tabular}
\end{table}

\subsection{Preprocessing module}
\label{sec:preprocessing_results}

We remove unmeaningful elements as detailed in Section \ref{sec:preprocessing} using the regular expressions in Listing \ref{lst:regular_expressions}. To remove the stop words, we use the \textsc{nltk} library\footnote{Available at \url{https://www.nltk.org}, October 2024. The stop words list was modified to exclude the terms `no', `yes', `more', `but', `very', `without', `much', `little', and `nothing'.}, and the content was lemmatized with the \texttt{en\_core\_web\_md} core model from the \texttt{spaCy}\footnote{Available at \url{https://spacy.io/models/en}, October 2024.} library.

\begin{lstlisting}[frame=single,caption={Regular expressions for preprocessing.}, label={lst:regular_expressions},emphstyle=\textbf,escapechar=ä]
ä\textbf{Numbers}ä = rä`ä\d+[A-Za-z]*'
ä\textbf{Punctuation characters}ä = rä`ä[\,ä$\mid$ä\.ä$\mid$ä'ä$\mid$ä:ä$\mid$ä;ä$\mid$ä\-ä$\mid-$ä]+'
ä\textbf{\textsc{url}s}ä = rä`ä(?:(pic.ä$\mid$ähttpä$\mid$äwwwä$\mid$ä\w+)?\:(//)*)\S+'
ä\textbf{Spaces}ä = rä`ä(\s|\t|\\n|\n)+'
ä\textbf{Special chararacters}ä = rä`ä(\*ä$\mid$ä\[ä$\mid$ä\]ä$\mid$ä=ä$\mid$ä\(ä$\mid$ä\)ä$\mid$ää\textbackslash\$ää$\mid$ä\"ä$\mid$ä\}ä$\mid$ä\{ä$\mid$ä\ä$\mid$ää$\mid$ä\+ä$\mid$ä&ä$\mid$ää\texteuroää$\mid$ä
ä\poundsää$\mid$ä/ä$\mid$ää\textdegreeä)+'
ä\textbf{Two-letter words}ä = rä`ä\b[a-zA-Z\-]{2,}\b
\end{lstlisting}

\subsection{Feature engineering module}
\label{sec:feature_engineering_results}

To generate features with \textsc{nlp}, the following libraries were used:
\begin{itemize}
 \item \texttt{textstat}\footnote{Available at \url{https://pypi.org/project/textstat}, October 2024.} to generate features 8, 10, 11, 14, 15 in Table \ref{tab:features}.
 
 \item \texttt{text2emotion}\footnote{Available at \url{https://pypi.org/project/text2emotion}, October 2024.} to generate feature 9.
 
 \item \texttt{spaCy}\footnote{Available at \url{https://spacy.io}, October 2024.} to generate features 12 and 13.

 \item \texttt{\textsc{gpt}-4o-mini} model of Chat\textsc{gpt}\footnote{Available at \url{https://platform.openai.com/docs/models/gpt-4o-mini}, October 2024.} to generate features 1-7. The system sends requests to Open\textsc{ai} \textsc{api}\footnote{Available at \url{https://openai.com/api}, October 2024.} with the prompt described in Listing \ref{lst:prompt_gpt} with the temperature parameter with value 0\footnote{This prevent the model to generate random responses for the same input.}.

 \item \texttt{CountVectorizer}\footnote{Available at \url{https://scikit-learn.org/stable/modules/generated/sklearn.feature_extraction.text.CountVectorizer.html}, October 2024.} from the \texttt{scikit-learn} Python library to generate feature 16. This library creates a vector of word occurrence frequencies using \SI{10}{\percent} of the data (approximately the first \num{1500} samples in our experimental dataset). The configuration parameters are as follows: {\tt ngrams\_max\_df} = 0.7, {\tt ngrams\_min\_df} = 0.01, and {\tt ngrams\_ngram\_range} = (1, 1). These values have been selected from experimental tests to minimize computational load while at the same time not compromising performance in the streaming setting.
\end{itemize}

\begin{lstlisting}[frame=single,caption={Feature engineering Chat\textsc{gpt} prompt.}, label={lst:prompt_gpt},emphstyle=\textbf,escapechar=ä]

This text has been taken from social media. Please 
analyze if it contains derogatory, humiliating, 
racist, sarcastic, sexual, threatening, or violent 
remarks, terms, and language, and return 1 if there 
exist and 0 otherwise. Following this JSON format. 
Do not add any explanation:
{"derogatory":0,
"humiliating":0,
"racist":0,
"sarcasm":0,
"sexual":0,
"threatening":0,
"violence":0
}
<Post>
\end{lstlisting}

\subsection{Feature analysis \& selection module}
\label{sec:feature_analysis_selection_results}

In order to reduce the number of features in the streaming system, we use the \texttt{SelectFromModel}\footnote{Available at \url{https://scikit-learn.org/stable/modules/generated/sklearn.feature_selection.SelectFromModel.html}, October 2024.} from the \texttt{scikit-learn} library in combination with the \textsc{rf} classifier\footnote{Available at \url{https://scikit-learn.org/stable/modules/generated/sklearn.ensemble.RandomForestClassifier.html}, October 2024.} as the meta-transformer wrapper selector. This analysis is performed during the cold start (\num{1500} first samples), resulting in \SI{36}{\percent} and \SI{18}{\percent} the features selected from the original set, in scenarios 1 and 2, respectively. This process is performed once at the experiment's beginning.

After a cold start, the features are evaluated by the \texttt{VarianceThreshold}\footnote{Available at \url{https://riverml.xyz/0.11.1/api/feature-selection/VarianceThreshold}, October 2024.} method. This function analyzes the variance variability and discards those features whose variance is 0. This process allows the selection of the most relevant features among the set derived from the first selection process during the cold start. Note that the set of selected features varies based on the variance value during the streaming life cycle.

\subsection{Classification module}
\label{sec:classification_results}

Our approach analyses the performance of the \textsc{gnb}\footnote{Available at \url{https://riverml.xyz/dev/api/naive-bayes/GaussianNB}, October 2024.}, \textsc{hatc}\footnote{Available at \url{https://riverml.xyz/0.11.1/api/tree/HoeffdingAdaptiveTreeClassifier}, October 2024.} and \textsc{arfc}\footnote{Available at \url{https://riverml.xyz/0.11.1/api/ensemble/AdaptiveRandomForestClassifier}, October 2024.} \textsc{ml} models in streaming.

The hyperparameters of each classifier were calculated in the cold start stage by applying an ad-hoc approximation of the \texttt{GridSearchCV}\footnote{Available at \url{https://scikit-learn.org/stable/modules/generated/sklearn.model_selection.GridSearchCV.html}, October 2024.} method for steaming operation. This method evaluates \textsc{hatc} and \textsc{arfc}\footnote{\textsc{gnb} has no hyperparameters to be tuned.} models in a range of parameters defined in Listings \ref{hatc_conf} and \ref{arfc_conf}, the selected parameters are marked in bold.

\begin{lstlisting}[frame=single,caption={\textsc{hatc} hyperparameter set.},label={hatc_conf},emphstyle=\textbf,escapechar=ä]
depth = [None, 50, 100, ä\bf200ä]
tiethreshold = [0.9, 0.5, 0.05, ä\bf0.005ä]
maxsize = [15, 50, 100, ä\bf200ä]
\end{lstlisting}
 
\begin{lstlisting}[frame=single,caption={\textsc{arfc} hyperparameter set.},label={arfc_conf},emphstyle=\textbf,escapechar=ä]
n_models = [10, 25, 50, 75, ä\bf100ä]
max_features = [sqrt, 4, 10, ä\bf25ä]
lambda_value = [2, 6, 10, ä\bf25ä]
\end{lstlisting}

Table \ref{tab:classification_results} shows the results for scenarios 1 and 2. \textsc{arfc} has the best performance, exceeding \SI{80}{\percent} on all metrics in scenario 1 and close to \SI{90}{\percent} in most metrics in scenario 2. More in detail, the difference in the cyberbullying recall metric between \textsc{arfc} compared to \textsc{gbn} and \textsc{hatc} is \SI{5}{\percent} and \SI{7}{\percent}, respectively. Regarding scenario 2, both \textsc{hatc} and \textsc{arfc} models surpass the \SI{80}{\percent} threshold in all metrics. However, the differences among the models are even more significant in this case, up to +\SI{17}{\percent} attained by \textsc{arfc} compared to \textsc{gnb} in the cyberbullying absent recall metric. Provided that the \textsc{arfc} is the best model, the difference in performance between the two scenarios is about \SI{5}{\percent}, which motivates us to discard content-derived features and is endorsed by the literature analysis.

\begin{table*}[!htbp]
\centering
\caption{\label{tab:classification_results}Classification results.}
\begin{tabular}{cccccccccccc}
\toprule
\multirow{2}{*}{\bf Scenario} & \multirow{2}{*}{\bf Model} & \multirow{2}{*}{\bf Accuracy} & \multicolumn{3}{c}{\bf Precision} & \multicolumn{3}{c}{\bf Recall} & \multicolumn{3}{c}{\bf \textsc{f}-measure} \\
\cmidrule(lr){4-7}
\cmidrule(lr){7-9}
\cmidrule(lr){10-12}
 & & & Macro & Absent & Present & Macro & Absent & Present& Macro & Absent & Present\\
\midrule

\multirow{3}{*}{1} 
& \textsc{gnb} & 83.03 & 83.67 & 79.25 & 88.10 & 82.97 & 89.91 & 76.04 & 82.93 & 84.24 & 81.63 \\
& \textsc{hatc} & 84.01 & \bf 85.37 & 78.67 & \bf 92.06 & 83.93 & \bf 93.72 & 74.14 & 83.83 & 85.54 & 82.13 \\
& \textsc{arfc} & \bf 85.00 & 85.20 & \bf 82.74 & 87.67 & \bf 84.97 & 88.78 & \bf 81.15 & \bf 84.97 & \bf 85.66 & \bf 84.28 \\

\midrule

\multirow{3}{*}{2} 
& \textsc{gnb} & 84.46 & 85.18 & \bf 90.44 & 79.92 & 84.52 & 77.37 & \bf 91.67 & 84.40 & 83.40 & 85.40\\
& \textsc{hatc} & 85.77 & 86.06 & 83.03 & 89.09 & 85.73 & 90.23 & 81.23 & 85.73 & 86.48 & 84.98 \\
& \textsc{arfc} & \bf 90.55 & \bf 90.84 & 87.63 & \bf 94.06 &\bf 90.52 & \bf 94.63 & 86.40 & \bf90.53 & \bf91.00 & \bf 90.06 \\

\bottomrule
\end{tabular}
\end{table*}

Our results offer a considerable improvement in comparison with the solutions from the state of the art selected in Table \ref{tab:comparison}: +\SI{41}{\percent} in cyberbullying recall than M.S. Islam and R. I. Rafiq (2023) \cite{islam2023comparative}, +\SI{22}{\percent} in cyberbullying precision than A. Sadek et al. (2023) \cite{Sadek2023}, and +\SI{11}{\percent} in the macro recall than E. A. Nina-Gutiérrez et al. (2024) \cite{nina2024multilingual}.

Regarding the most closely related works, H. Herodotou et al. (2020) \cite{Herodotou2020} proposed a streaming \textsc{ml} system without explainability. Note that their approach relies exclusively on content-dependent features. In this regard, better performance is attained without generalization and bias limitations. Moreover, the convergence of our system is faster (+\SI{39}{\percent} macro \textit{F}-measure with the \num{4000} initial samples). Ultimately, Y. Yan \textit{et al.} (2023) \cite{Yan2023} is the only proposal combining \textsc{ml} and \textsc{llm}s, however in batch setting and without explainability capabilities. Our results improve the performance of the latter solution in +\SI{13}{\percent} for the macro precision metric.

\subsection{Explainability module}
\label{sec:explainability_results}

\begin{figure}[!htpb]
 \centering
 \includegraphics[scale=0.1]{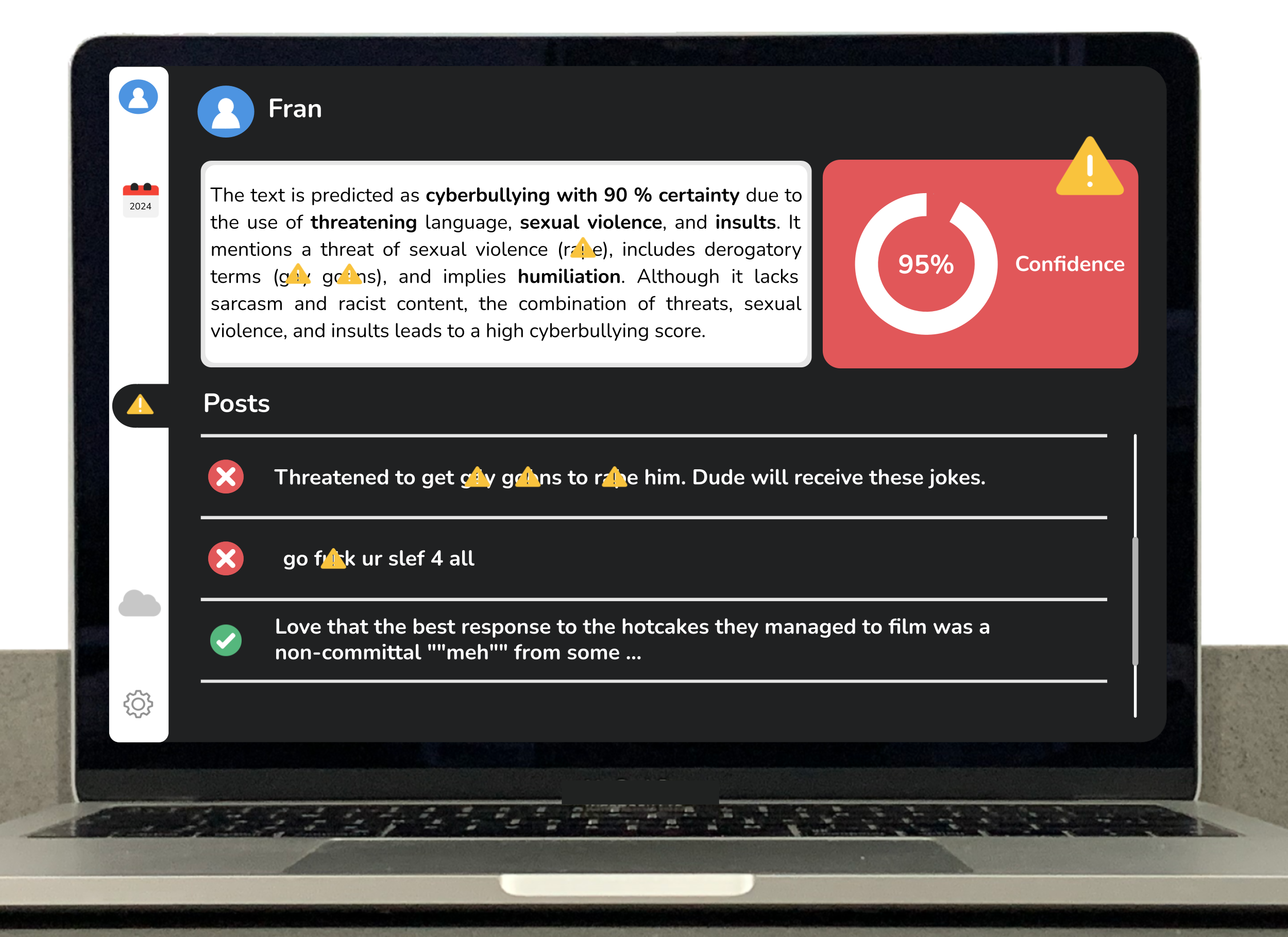}
 \caption{Dashboard for explainability.}
 \label{fig:dashboard}
\end{figure}

We apply the method described in Section \ref{sec:explainability} to generate natural language explanations of the \textsc{arfc} model predictions. Listing \ref{lst:prompt_gpt_expl} shows the prompt template used to guide the \texttt{\textsc{gpt}-4o-mini} model using the Open\textsc{ai} \textsc{api}. The template includes the predicted category, the confidence percentage by \texttt{\small Predict\_Proba\_One} function\footnote{Available at \url{https://riverml.xyz/0.11.1/api/base/Classifier}, October 2024.}, the feature values, and the post content. Figure \ref{fig:dashboard} shows a dashboard with the list of collected posts and the expected category (see bottom). The current description generated by the \textsc{llm} is placed at the top of the screen, and on the right, the category predicted (presence of cyberbullying in red and its absence in green) along with the confidence interval. 

\section{Conclusion}
\label{sec:conclusion}

The proliferation of new social media networks and platforms results from the growing use of modern society for virtual interaction and communication. Even though they provide instant and ubiquitous connectivity, they also promote unlimited access to other members that can encompass negative behavior like sharing sensitive content, sending threatening messages, or spreading false information and rumors. Among the negative consequences, cyberbullying deserves attention since it has a direct detrimental impact on people's health.

\begin{lstlisting}[frame=single,caption={Chat\textsc{gpt} prompt for explainability.}, label={lst:prompt_gpt_expl},emphstyle=\textbf,escapechar=ä]

Generate an explanation in less than 500 characters 
that indicates why this post was categorized as 
[cyberbullying / no-cyberbullying] with [xx%]. Note
that the features used by the model that generated 
the prediction are: derogatory=[0/1], 
humiliation=[0/1],racist=[0/1], sarcasm=[0/1], 
sexual=[0/1],threatening=[0/1] and violence=[0/1].
<Post content>

\end{lstlisting}

We have detected a significant gap in the cyberbullying literature regarding real-time solutions (\textit{i.e.}, those that operate in streaming). Moreover, scant research has been performed on jointly leveraging \textsc{llm}s language understanding capabilities and using highly accurate detection approaches like \textsc{ml}. Ultimately, the lack of interpretable and explainable solutions deserves mention. Accordingly, we propose an innovative and real-time solution for cyberbullying detection that combines stream-based \textsc{ml} and \textsc{llm}s. The contributions of our work are complete with an explainable dashboard that provides the rationale behind the classification prediction and its confidence.

Results on experimental data report promising performance close to \SI{90}{\percent} in all evaluation metrics and surpassing those obtained by competing works in the literature. Ultimately, our proposal promotes trustworthiness, reliability, accountability, and online safety, which contributes to palliating the negative consequences of cyberbullying. In future work, we plan to follow a multi-label approach to tag comments on different abusive behavior (\textit{e.g.}, racist, sexist) and apply a sliding-window-based feature engineering, analysis, and selection to consider past user behavior. Moreover, we would like to perform experiments on non-human generated data to fight the proliferation of hate bots in online communities. Multi-modal data (\textit{e.g.}, images) will also be considered. The ultimate objective is to analyze the long-term adaptability and performance of the proposed online solution to provide personalized and targeted interventions and its results regarding bias and discrimination vulnerability.

\section*{Declaration of competing interest}

The authors have no competing interests to declare relevant to this article's content.

\section*{Acknowledgment}

This work was partially supported by Xunta de Galicia grants ED481B-2022-093 and ED481D 2024/014, Spain.

\bibliography{2_bibliography.bib}
\bibliographystyle{IEEEtran}
\end{document}